# Stability prediction of the software requirements specification

José del Sagrado, Isabel M. del Águila

Department of Informatics, University of Almería, 04120 Almería, Spain

**Abstract**
Complex decision-making is a prominent aspect of Requirements Engineering. This work presents the Bayesian network Requisites that predicts whether the requirements specification documents have to be revised. We show how to validate Requisites by means of metrics obtained from a large complex software project. Besides, this Bayesian network has been integrated into a software tool by defining a communication interface inside a multilayer architecture to add this a new decision making functionality. It provides requirements engineers a way of exploring the software requirement specification by combining requirement metrics and the probability values estimated by the Bayesian network.

## 1 Introduction

Since the appearance of the first intelligent editors, the challenge of supporting software development using Artificial Intelligence techniques last for over 20 years. In spite of the success of some punctual results and progresses made during these years, the intelligent environment for software development is still considered under construction, perhaps because software engineers are typically focussed on prosaic and practical engineering concerns rather than building smart algorithms (Harman, 2012).

Expert knowledge is involved in every software development project, since developers must face decisions based on their expertise during all the development stages from requirements to maintenance (Meziane and Vadera, 2010). That is to say, Software Engineering (SE) can be considered as a knowledge intensive process and it can be framed within the AI domain (Harman, 2012). The software engineering community has used many algorithms, methods, and techniques that have emerged from the Artificial Intelligence community (Shirabad and Menzies, 2005; Zhang et al., 2012).

Furthermore, if a portion of expert knowledge could be modeled and incorporated in the software engineering lifecycle, as well as in the tools that support this lifecycle, we would obtain a great benefit for any development process.

This work presents how the Bayesian network model, which allows the assessment of Software Requirements Specification (SRS), called Requisites (del Sagrado and del Águila, 2010) has been validated using data measured from a specific software development project dataset and how it has been embedded in a previously built software tool. The model was designed to be used as a predictor that tells us whether the software requirements specification (SRS) has enough quality, stopping the iterations needed in order to define the SRS. Once measures were computed or obtained, their values will be introduced as evidence. Thus, the propagation of belief will be used to determine if the requirements specification have the enough quality to define a baseline for the project. On the other hand, we present a successful solution that instantiates an architecture for the seamless integration of a CARE (Computer Aided Requirement Engineering) tool in order to manage requirements with some AI techniques (del Sagrado et al., 2011). Specifically, we present the integration in an academic CARE tool, InSCo-Requisite (Orellana et al., 2008), with the Bayesian network. This tool has been extended with a new functionality that makes the adaptation between the requirements metrics and the variables used in the Bayesian network, allowing the use of this AI technique in the requirements specification stage.

The rest of the paper is structured as follows. After describing the basic requirements workflow, section 2 describes the related works about the benefit of using AI in software development. Section 3 includes de description of the Bayesian network Requisites and how it has been validated using a real world large scale dataset. Section 3 is devoted to the task of define the process followed to the integration of Requisites in the tool InSCo-Requisite in order to define a software project baseline. In this section, some examples of use a specific software development project are also given. The paper ends with conclusions in section 5.

## 1.1 Background. Requirements workflow

Requirements express the needs and constraints established for a software product that contribute to the solution of some real world problem (Kotonya and Sommerville, 1998). Requirements development is considered a good domain for the application of AI techniques, because requirements by themselves tend to be imprecise, incomplete and ambiguous. This area of SE is quite different from others because requirements belong to problem space, whereas other artefacts, obtained in software development, reside in the solution space (Cheng and Atlee, 2007). If requirement-related tasks are poorly executed, usually the software product obtained becomes unsatisfactory from a software factory point of view (Sommerville, 2011; Standish Group, 2008).

Requirements are critical to the success of a software project, because they collect the needs or conditions that have to be met by the product. That is, they are the basis for the rest of the development process. Therefore, any improvement that takes place during the requirements development stage will favourably affect the whole production life-cycle. This stage is articulated by the execu-tion of several activities and has been defined, by different authors (Kotonya and Sommerville, 1998; Abran et al., 2004; Wiegers and Beatty, 2013), as a process with variations. The simplest set of activities for creating requirements (Alexander and Beus-Dukic, 2009) comprises discovery, documentation and validation. The discovering-documenting-validating cycle (DDV) is carried out through several iterations in order to complete the requirements specification and then moves towards the next development task.

*Discovering requirements* is the task which determines, through communication with customers and users, what are their requirements. Requirements are elicited (or gathered) through interviews and other techniques such as stakeholders workshops or inspections. In this first activity is where the problem, that software is going to solve, has to be understood. It is usually a complex task, because this activity requires good communication between software users and software engineers. Requirements have to be conceived without ambiguities in order to define what the system is expected to do.

*Documenting requirements* is about capturing software requirements. These requirements are captured in a document or its electronic equivalent, known as Software Requirements Specification (SRS). Software requirements documents play a crucial role in SE (Nicolás and Toval, 2009). Early approaches to perform this activity used to work with word processors, but this method of supporting SRS was prone to error and tedious. CARE (Computer-Aided Engineering Requirement) tools appeared to give a solution to these problems, providing environments that make use of databases, allowing an effective management of the requirements of any software project. These tools also allow the use of modelling languages in requirements description (i.e. use cases, UML) or informal languages (storyboards).

*Validating requirements* is in charge of checking if requirements present inconsistencies, ambiguities or errors. This task is concerned with the process of analysing requirements in order to detect or resolve conflicts, and properly define the bounds of the software system. The process of reaching a consensus on an appropriated trade-offs, i.e. requirements negotiation, is performed during this activity, as well. The need of a negotiation appears when two stakeholders require mutually incompatible features or when requirements require more development resources than those that are really available.

## 2 Related works

The existing works have already demonstrated that there is considerable potential for software engineers to benefit from AI. The algorithms, methods, and techniques emerged from the Artificial Intelligence community, which have been intertwined Software Engineering (SE) community can be arranged in three major areas: 'Search Based Software Engineering'(SBSE), 'Classification, Learning and Prediction for Software Engineering' and 'Probabilistic Software Engineering', (Harman, 2012). SBSE re-formulates SE problems as optimisation problems and it has proved to be a widely applicable and successful approach from requirements and design stages, a bibliometric analysis in of many publications of SBSE describes how this area have grown since 2001 (Harman et al., 2012). In classification, learning and prediction some authors propose models for the prediction about risky issues in SE (Menzies and Shepperd, 2012), either related to the study of the defects (Kastro and Bener, 2008), or the process of predicting the effort required to develop a software system (Wen et al., 2012). A probabilistic AI technique with a high applicability in SE is Bayesian probabilistic reasoning to model different software topics (Misirli and Bener, 2014), as quality management (Tosun et al., 2015) or defect prediction (Mısırlı et al., 2011). There is a blurred border between these three major areas, so some of these works can be included in more than one.

Nevertheless, focusing on requirements stage, the requirements engineering is the less covered by the AI approaches. SBSE focus on requirements only in a 3% of works, (Harman et al., 2012) (de Freitas and de Souza, 2011), and papers that deals about how to apply Bayesian networks to requirements are few (del Águila and del Sagrado, 2015). Artificial Intelligence can provide a new dimension to the stage of requirements development, by defining methods and tools for the engineer that allow a simpler execution of the entire software development project. The task of discovering requirements can be assisted by machine learning techniques that allow to organize collaboration between stakeholders using clustering techniques to manage discussion forums about requirements (Castro-Herrera et al., 2009) or that allow the automatic clustering of product features for a given domain (Dumitru et al., 2011). Requirements can be considered as the bricks gluing different stages in software project development. So, if we have a risky requirements process, probably we will have a risky project. In order to mitigate the risks, we need to identify and assess the risks of requirements. Bayesian networks classifiers can assists the process of predicting the risk level of a given requirement (del Águila and del Sagrado, 2011). Resource constraints usually appear in earlier development stages and prevent the development of all defined requirements, forcing developers to negotiate requirements. Therefore, a basic action is the selection of the set of requirement to be included in the next steps of the development project. This problem, known as the next release problem (Bagnall et al., 2001), has received attention by AI researchers and it is considered as an optimization problem (Bagnall et al., 2001; Karlsson and Ryan, 1997; Greer and Ruhe, 2004; del Sagrado et al., 2015).

Probabilistic approaches have less used in RE, likely because RE decision making is not sufficiently mature, and no closed set of decision problems in RE are available that could be tackled with probabilistic models. Furthermore, there are several major challenges about how to apply BN to RE, such as how to deal with networks validation, or how to embed the models obtained in computer-aided software engineering tools (del Águila and del Sagrado, 2015). That is the reason we include in this paper not only the probabilistic model and how it has been validated but also the integration of the model in an academic CARE tool InSCo-Requisite (Orellana et al., 2008) that has been extended with a new functionality.

# 3 A Probabilistic Requirements Engineering solution

Bayesian networks (BNs) have been used for decision making in SE for many years. Bayesians networks (Jensen, 2007; Kjaerulff and Madsen, 2007) allow us to represent graphically and concisely knowledge about an uncertain domain. A Bayesian network has:

- a qualitative component, $G = (U, A)$, which is a directed acyclic graph (DAG), where the set of nodes, $U = \{V_1, V_2, \cdots, V_n\}$, represents the system variables, and the set of directed edges, $A$, represents the existence of dependences between variables

- a quantitative component, $P$, which is a joint probability distribution over $U$ that can be factorized according to:

$$P(V_1, V_2, \cdots, V_n) = \Pi_{i=1}^{n} P(V_i/Pa(V_i)) \qquad (1)$$

where $P(V_i/Pa(V_i))$ is the conditional probability for each variable $V_i$ in $U$ given its set of parents $Pa(V_i)$ in the DAG.

The structure of the associated DAG determines the dependence and independence relationships among the variables. Besides the local conditional probability distributions measure the strength of the direct connections between variables.

A BN can be used as a predictor simply by considering one of the variables as the class and the others as features that describe the object that has to be classified. The posterior probability of the class is computed given the features observed. The value assigned to the class is the one that reaches the highest posterior probability value. A predictor based on a BN model provides more benefits, in terms of decision support, than traditional predictors, because it can perform powerful what-if problem analyses.

## 3.1 Bayesian Network Requisites

BNs are very useful in SE, since its representation of causal relationships among variables is meaningful to software practitioners (Harman, 2012), (Misirli and Bener, 2014). A specific case of use of this AI technique in requirements workflow is the Bayesian network Requisites (del Sagrado and del Águila, 2010). It has been built, through interaction with experts and using several information sources, such as standards and reports. Its aim is to provide developers an aid, under the form of a probabilistic advice, helping them at the time of making a decision about the stability of the current requirements specification. Requisites provides an estimation of the degree of revision for a given requirements specification (i. e. SRS). Thus, it helps the process of identifying if a requirements specification is stable and does not need further revision (i.e. if it is necessary or not to perform a new DDV cycle).

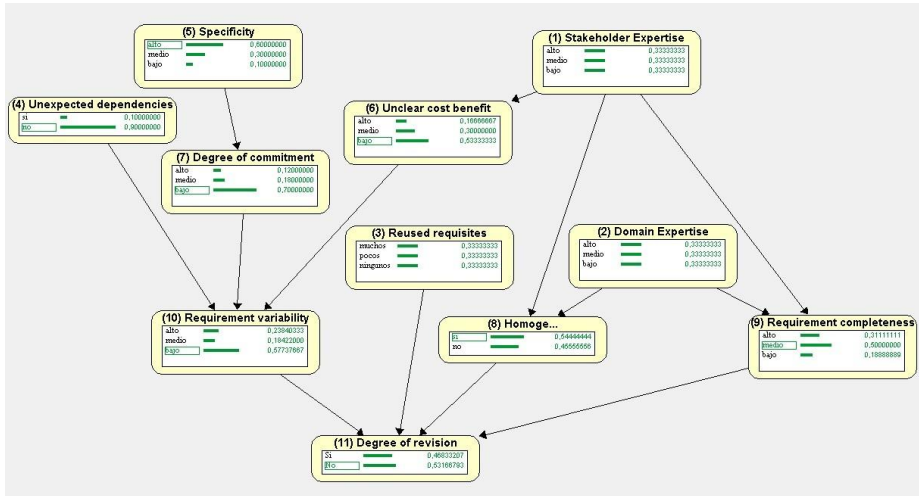

Figure 1: Bayesian Network Requisites.

The structure of Requisites (see Figure 1) contains variables (see Table 1) and dependencies in order to assess the goodness of SRS. Once the structure was established, the probability values, that define the quantitative part of the network, were fixed with the aid of two software engineers in an interview-evaluation cycle. Once the values have been computed, they will input as evidence and the propagation of this belief over the network will be used to determine whether the requirements specification should or should not be revised.

It is worth noting some of the relationships between variables that occur in the network. Thus, if the 'degree of commitment' (i.e. the number of requirements that have to be agreed) increases, then the level of 'specificity' will drop. If stakeholders have little 'experience' in the processes of requirements engineering, it is more likely to obtain requirements which are 'unclear' in terms of cost/benefits. The 'requirement completeness' and the 'homogeneity of the description' are influenced by the 'experience of software engineers' in the domain of the project and by the 'stakeholders expertise' in the processes or tasks of requirements engineering. If experience is high, the specification will be 'complete and homogeneous' because developers have been able to describe the requirements with the same level of detail all requirements they have been discovered. Finally, the 'requirement variability' represents the number of changing requirements. A change in the requirements will be more likely to occur if 'unexpected dependencies' are discovered or if there are requirements that do not add any value to the end software or if there are missing requirements or if requirements have to be negotiated.

There are free software packages for the construction and use of Bayesian networks. In this work, we used Elvira (Elvira Consortium, 2002), a package that allows the implementation of a Bayesian Network model using Java classes to support the model itself and the inference process needed to use it. However, from a practical point of view, the integration of a built network within a software system is not trivial, because it is necessary to define the communication paths between both components by matching variables and results.

Table 1: Variables in Requisites (del Sagrado and del Águila, 2010)

| Variable | Description | Value |
|---|---|---|
| Stakeholders´ expertise | Represents the degree of familiarity that stakeholders have in respect to Requirements Engineering processes. If stakeholders have already collaborated in other projects, using techniques and skills of the Requirements Engineering discipline they will play a clearer role and will commit fewer errors. | High, Medium, Low |
| Domain expertise | Expresses the level of knowledge that the development team has about the project domain. When developers use the same concepts than stakeholders, communication will be correct and it will be required a smaller number of iterations | High, Medium, Low |
| Reused requirement | Checks if there are reused requirements. The reuse tries to reduce the development cost by enhancing the productivity of the development team. Thus, if the number of requirements that come from reusable libraries is high, in general, specification of the requirements does not need new iterations. | Many, Few, None |
| Unexpected dependencies | In some cases, unexpected dependencies or relationships can appear between requirements or groups of them. This fact usually involves a new revision of the specification of the requirements. | Yes, Non |
| Specificity | Represents the number of requirements that have the same meaning for all stakeholders. That is, if stakeholders use the same semantic, we will need less revision and a shorter process of negotiation in order to reach a commitment. | High, Medium, Low |
| Unclear cost/benefit | Represents that stakeholders or developers include requirements that do not have direct quantifiable benefits for the business or the organization in which the software to be developed will operate. | High, Medium, Low |
| Degree of commitment | Represents the number of requirements that have needed a negotiation to be accepted. The requirements of a project are a complex combination of requirements from different stakeholders, and some of them can generate conflicts that unbalance the specification. | High, Medium, Low |
| Homogeneity of the description | A good SRS must be described at the same level of detail. If some requirements have been described in a detailed way, all the requirements of this SRS should be described at the same level of detail. If there is not homogeneity, the SRS will need to be revised. | Yes, Non |
| Requirement completeness | Indicates if all significant requirements have been elicited and/or specifie7d. | High, Medium, Low |
| Requirement variability | Represents that requirements have suffered changes. If the specification of a requirement changes, it is quite possible that this modification will affects the whole SRS, and an additional revision is likely needed. | High, Medium, Low |
| Degree of | Is the value predicted by Requisites and it indicates | Yes, |

## 3.2 Validation of Requisites

The Bayesian network Requisites makes a prediction indicating whether a requirements specification is sufficiently accurate or requires further revision. The inference process, that uses the evidences offered by the metrics calculated over the current version on the SRS, to calculate the marginal probability distribution of the variable 'degree of revision'.

A real world large scale dataset is adopted to evaluate the approach of Requisites. RALIC is the acronym for Replacement Access, Library and ID Cars and it was a large-scale software project to replace the existing access control system at University College London and consolidate the new system with library access and borrowing (Lim and Finkelstein, 2012). The objectives of RALIC include replacing existing access card readers, printing reliable access cards, managing cardholders information, providing access control, and automating the provision and suspension of access and library borrowing rights. The stakeholders involved in the project had different and sometimes conflicting requirements. The project duration was 2.5 years, and the system has already been deployed at University College London. We have obtained the evidence values measuring this dataset and these metrics are injected to the BN to analize the inference through the probabilistic model.

RALIC requirements were organised into three levels of hierarchy: project objectives, features, and specific requirements. A feature that contributed towards a project objective was placed under the project objective, and a specific requirement that contributed towards the feature was placed under the feature. These requirements organization indicates the level of detail at which requirements are described, i.e. the homogeneity of the description. In order to set the homogeneity of the description value, we can study the level of detail reached by each of the project objectives. We study the proportion of the requirements linked to a project objective that have been described in terms of specific requirements. Figure 2 shows in a box-plot the distribution of the percentage of detail applied to describe project objectives. The 75 percent of project objectives have been described in terms of specific requirements in a percentage above 50.96. This indicates that the branches in the hierarchical structure of the project have a similar depth, which translates into an homogeneous description. Thus, the value of Homogeneity of description is set to 'yes' in Requisites.

Stakeholders are asked to rate requirements rather than rank all of them, because previous work has shown that large projects can have hundreds of requirements, and stakeholders experienced difficulty providing a rank order of requirements when there are many requirements (Lim and Finkelstein, 2012). Each stakeholder assigns ratings to the set of requirements identified by the project team. A rating is a number on an ordinal scale (e.g., 0 – 5) reflecting the importance of the requirement to the stakeholder (e.g., 0 means that the requirement is not considered by the stakeholder; 1 means that the requirement is not very important to the stakeholder; 5 means that the requirement is very important).

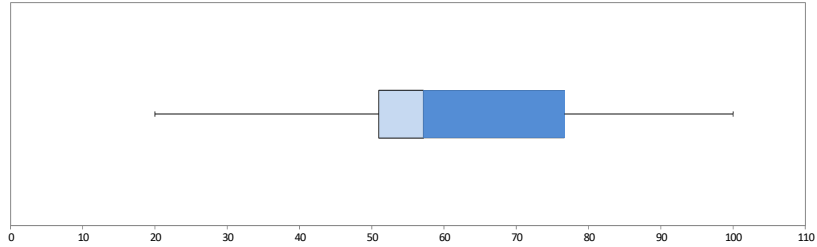

Figure 2: Distribution of the percentage of detail applied to describe project objectives.

Specificity has to deal with the meaning of requirements for stakeholders. So, the more number of stakeholders agree on the meaning of requirements, the lesser revision we need. In order to measure the specificity value, we used a consensus measure, the average of the ratings assigned to a given project objective. That is to say, the higher the average rating of a project objective is, the higher specificity we have. Figure 4.a shows the distribution of the average rating of each project objective, whilst figure 4.b shows the distribution of the specificity of each project objective. The specificity value of each project objective has been computed directly from its average rating by adapting the range from $\{0, 1, 2, 3, 4, 5\}$ to $\{1(low), 2(medium), 3(high)\}$. As result of these process the value of specificity will have to set to 'high' in Requisites, because the 90% of the project objectives map to an specificity value of 'high'.

The stakeholder priority data for RALIC was also collected (Lim and Finkelstein, 2012). The stakeholders were asked to recommend people whom they think are stakeholders in the project. Their recommendations were then used to build a social network, where the stakeholders were nodes and their recommendations were links. The output was a prioritized list of stakeholders and their requirements preferences.

Stakeholders were requested, through the OpenR questionnaire, to make recommendations on the salience (i.e. level of influence on the system) of other stakeholders. The salience of a stakeholder is assigned as an ordinal variable whose domain is the set $\{1, 2, 3, 4, 5, 6, 7, 8\}$. Salience an expertise are in straight relationship, as the influence on a system an Stakeholder has, can be considered also as a measure of its level of expertise. In order to get an overall estimation of the stakeholders' expertise, first we have resumed the recommendations received by each stakeholder as the average (Figure 5.a shows the stakeholders' salience distribution computed in this way). Then the expertise of each stakeholder is obtained mapping her/his salience to the set $\{1(low), 2(medium), 3(high)\}$. Figure 5.b shows the stakeholders' expertise distribution computed in this way. It can be observed that 92% of the stakeholders receive a 'low' expertise recommendation, thus the value of stakeholders' expertise is set to 'low' in Requisites.

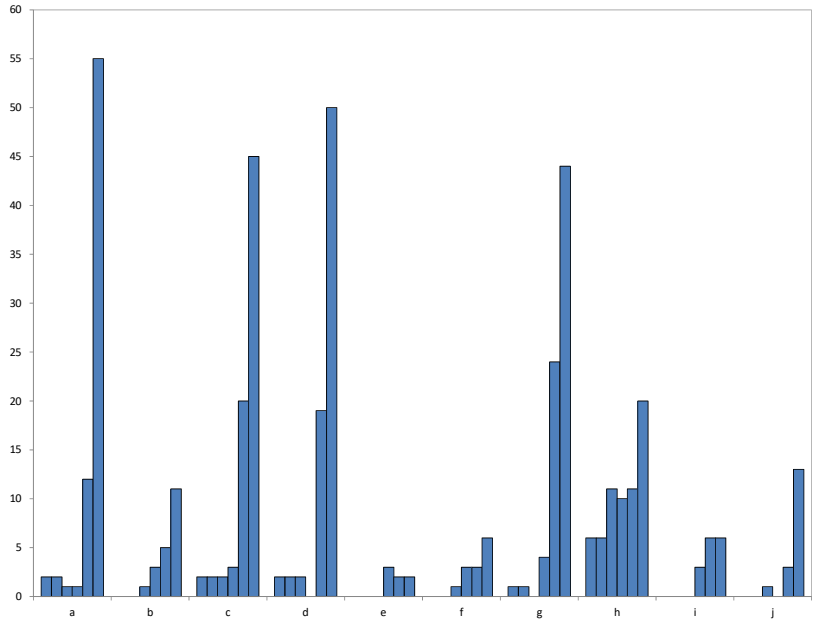

Figure 3: Ratings assigned by stakeholders to project objectives.

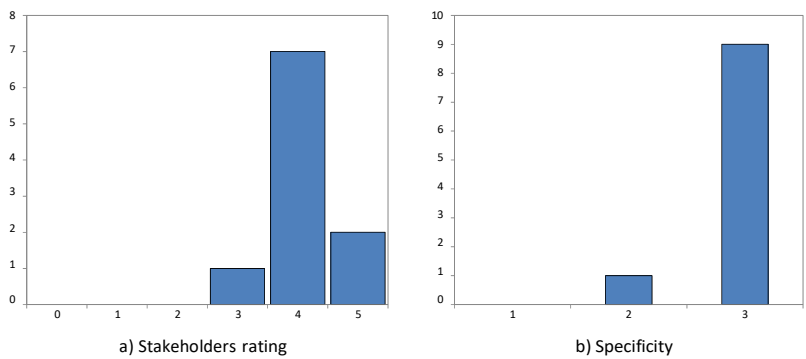

Figure 4: Distribution of project objectives' rating and specificity.

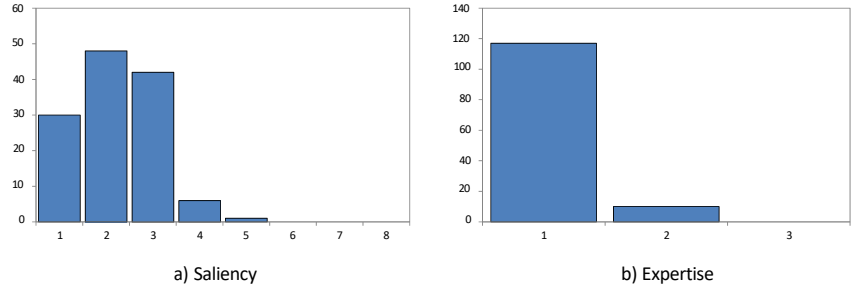

Figure 5: Distribution of stakeholders' salience and expertise.

Then, we need to study the behaviour of the BN Requisites when the data extracted from RALIC dataset are incorporated to the model. The a priori probabilities are shown in Figure 1. If the value of the variable 'homogeneity of the description' is set to 'yes' the probability of the value 'no' to the degree of revision will arise to 0.54. That is, the more uniform is the overall description of the requirements, the less revision is needed. This trend is reinforced when we included the evidence value for 'specificity', the 'degree of revision' gets 0.45, 0.55 for 'yes' and 'no' respectively. Nevertheless, because of the 'expertise for stakeholders' reach a value of 'low' the final prediction, shown in Figure 6, is that we have to review the SRS. The final values are 0.52 for 'yes' and 0.48 for 'no'.

## 4 Integrating Requisites in a CARE tool

One of the bigger breakthroughs in requirements management workflow was produced when we stopped thinking of documents and started thinking about information. So, to be able to handle this information you have to resort to databases, specifically to documentary databases that have evolved into what nowadays are called CARE tools. There is a raising number of CARE tools that are currently available on the market (de Gea et al., 2012). Among them, the best known are IRqA (Visure Solutions, 2012), Telelogic DOORS (IBM, 2012) and Borland Caliber (Borland Software Corporation, 2016). InSCo-Requisite is an academic web CARE tool, developed by DKSE (Data Knowledge and Software Engineering) group at the University of Almería, which supports partially the requirements development stage (Orellana et al., 2008). It provides a basic functionality where groups of stakeholders collaboratively work through the Internet in order to define the SRS. Because of the opportunity to easily make changes in this tool, we have an exceptional chance to tackle the integration of AI techniques in a CARE tool. Besides, this solution successfully instantiates an architecture for the seamless integration of a CARE (Computer Aided Requirement Engineering) tool in order to manage requirements with some AI techniques (del Sagrado et al., 2011).

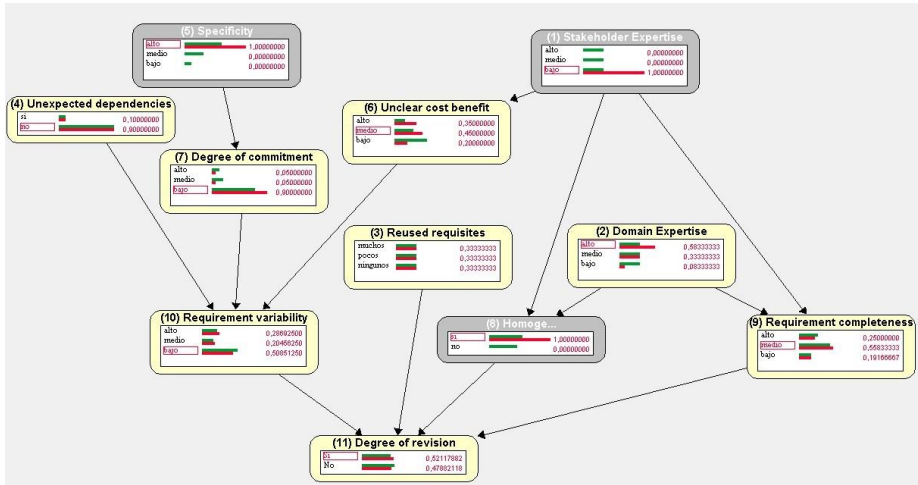

Figure 6: Requisites state after including the evidence obtained from RALIC.

## 4.1 InSCo-Requisite

Commercial CARE tools offer powerful solutions to capture the requirements of a software development project and also include methods to analyze the requirements or to monitor the changes on each requirement. The purpose of InSCo-Requisite is not to compete against these commercial tools. It has been developed within an academic setting to deal with the problem of the management of software projects that include components based and not based on knowledge (del Águila et al., 2010; Cañadas et al., 2009). Its main goal is to offer, under a distributed environment, an intuitive and easy-to-use tool to manage requirements.

Requirements are more than a list of 'the system shalls'. In a broad sense, requirements are a network of interrelated elements or artifacts. These artifacts (objectives, constraints, priorities, ...) must be modified and managed during the task of discovering requirements. Our tool guides requirements management by means of templates, which can be classified into two groups for *functional requirements*: *objectives*, which specify business-related information, and *user requirements*, which are related to customer needs. At the lowest level of refinement, templates are fulfilled using natural language and scenarios. *Non functional requirements* have their own template to collect the quality attributes and the constraints. Figure 7 shows the InSCo-Requisite artifact model. Users can be assigned to various projects, which are organized into folders that permit a hierarchical structuring of the requirements. A user can participate in a project by proposing new requirements, changes in requirements or comments on them (both own and from other).

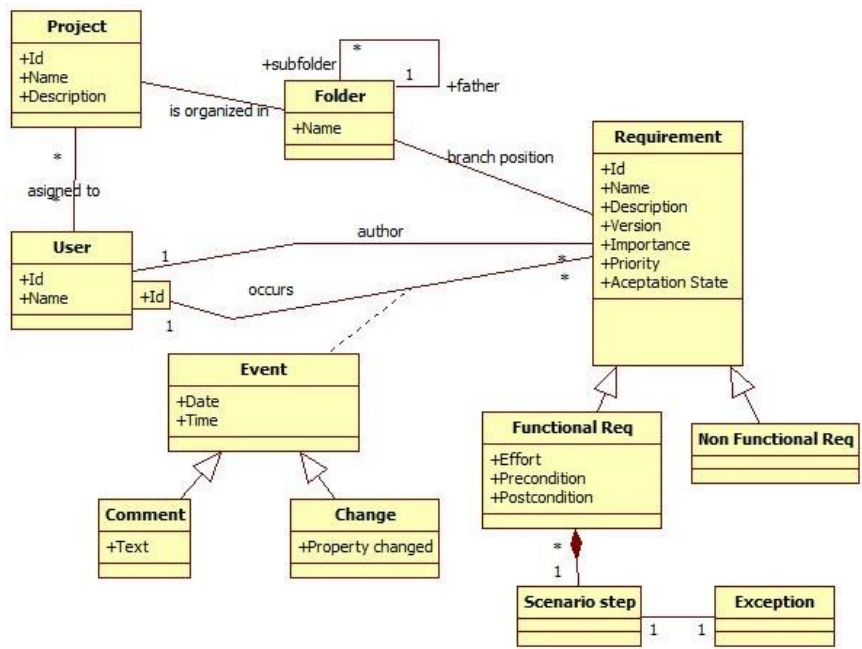

Figure 7: Conceptual model of InSCo-Requisite.

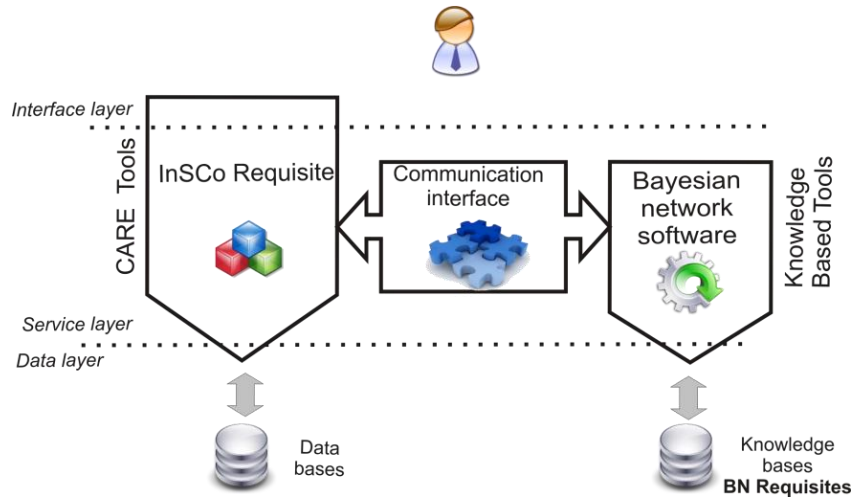

Figure 8: Generic Architecture.

We can represent relationships between requirements, as those defined between objectives and users' requirements or between. The tool maintains a hierarchical structure of templates, which shows a global perspective of the project content and displays parent/child relationships between templates in a explorer window.

InSCo-Requisite offers some facilities for informal collaboration through rich contextual discussions about requirements. Users can start discussions about any of the existing requirements in order to propose changes, improvements, or discuss about the template content. Therefore the tool provides access to a change log for every requirement, stakeholder or project. For any requirement, there is no limit in terms of events that can be assigned to it or the number of changes or comments that a stakeholder can make on it.

Java EE (Java Enterprise Edition) is the platform chosen for developing and deploying InSCo-Requisite. It has a multi-layered architecture based on Struts, an open-source multi platform framework created by the Apache Software Foundation. The architecture adopted in the first version of the tool separates presentation logic from business logic and persistent storage. The interface layer represents the client side, that is to say, web browsers that send requests to the application server at the service layer, using the HTTP (HyperText Transfer Protocol). This layer is also in charge of representing the data received from the service layer, basically under the form of HTML (HyperText Markup Language), CSS (Cascading Style Sheet) and Javascript code. The service layer is composed by an Apache Tomcat server that gives support to the Java EE Platform for the InSCo Requisite application. The server processes the incoming requests, executes the appropriated Java servlets, and exchanges data with the data layer by means of a JDBC (Java DataBase Connection). Finally, the data layer, is in charge of keeping the data persistence and is composed by an Oracle Database Server.

## 4.2 Connection between Requisites and InSCo-Requisite

It would be considerably helpful, for any development team, to have AI techniques (i.e. Bayesian network) available, as an aid, inside a CARE tool (del Sagrado et al., 2012). The BN Requisites aids in the requirements definition process during a specific software project, as we have shown in section 3.2. Since the requirements management task is performed by means of a CARE tool (InsCo-Requisite), this tool must provide the information required by the network (evidence) in order to obtain the value for the variable (e.g. degree of revision) that we want to predict.

But, AI techniques and CARE tools have been developed independently of each other. Therefore, it is necessary to define a communication interface between them, preserving the independent evolution of both areas and achieving a synergetic profit between them (del Sagrado et al., 2011). The CARE tool is in charge of the management of all the information related to the development project (requirements, customers, etc) which is stored in a database. Next, the communication interface connects the CARE tool and the Bayesian network, interchanging the required information needed for the execution of the appropriated processes and adapting the languages used by each tool (see figure 8). Thus, the task of validating requirements receives metrics from the SRS and returns an estimation of the degree of revision for the SRS, that is the knowledge-based tool helps the requirements validation task according the DDV cycle. All of these communication processes are performed through XML files.

The BN Requisites makes a prediction indicating whether a requirement specification is sufficiently accurate. The evidences (i.e. the observed variables) are provided by the CARE tool InSCo-Requisite, that is in charge of extracting the values of variables from: the data about projects, requirements, users' activity and so on. Table 2 shows how these evidences are obtained. For example, the 'stakeholders expertise' is estimated by analyzing all the projects in which the stakeholders of the current project have participated on, together with their degree of participation and the requirements that have been affected by this participation; the 'specificity' is estimated by counting the number of changes or comments made on requirements by two or more stakeholders. Observe that currently, some variables cannot be obtained from the InSCo-Requisite tool, and must be estimated by users, as e.g. 'domain expertise' or 'reused requirement', but our architecture allows to include these metrics in Requisites when InSCo-Requisite would be able to measure them. Once these values have been collected, it is necessary to combine and discretize them in order to establish a correspondence with the values of the variables in Requisites (see last column in Table 2).

Table 2: Connections between measures in InSCo-Requisite and Requisites

| Requisite Variables | Measures in InSCo-Requisite | |
|---|---|---|
| Stakeholders' expertise | For every stakeholder in the project counts: <br><br> Projects to which stakeholders have been assigned. <br> Requirements with stakeholders participation. | High, Medium, Low |
| Domain expertise | Manual assignment. | High, Medium, Low |
| Reused requirement | InSCo-Requisite does not support it. | Many, Few, None |
| Unexpected dependencies | InSCo-Requisite does not support it. | Yes, No |
| Specificity | Number of ACCEPTED requirements in which: <br> Several stakeholders have participated. | High, Medium, Low |
| Unclear cost/benefit | Number of requirements whose: <br><br> Status has changed (ACCEPTED-REJECTED) <br> Comments have been sent by several stakeholders. | High, Medium, Low |
| Degree of commitment | Number of requirements in which: <br><br> Several stakeholders have sent comments <br> Several stakeholders have performed changes | High, Medium, Low |
| Homogeneity of the description | All the branches in the hierarchical structure of the project have a similar depth. | Yes, No |
| Requirement completeness | Level of fulfilment of template fields of all the requirements. | High, Medium, Low |
| Requirement variability | Number of changes registered on requirements. | High, Medium, Low |

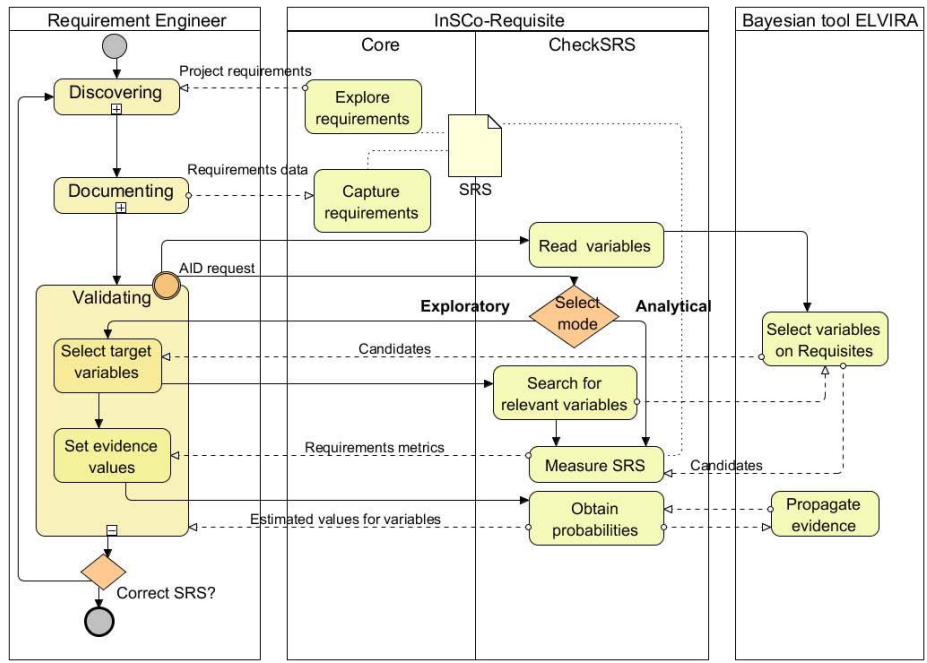

Figure 9: A processes model for Requirements Engineering

## 4.3 A Use Case

The requirements engineer executes several times the DDV cycle until gets a complete requirements specification. The whole process is shown in Figure 9, and the enhancement, obtained by the integration of Requisites in InSCo-Requisite, specifically concerns to the task of validating requirements.

When an aid request for validating requirements starts, InSCo-Requisite asks Requisites for its set of variables. At this point, requirements engineer chooses one of two modes: *analytic* or *exploratory*. The *analytic mode* (see Figure 10) is used when the user wants to obtain the posterior probability distribution of any target variable, given some evidence (i.e. an assignment of the values of some of the other variables). In *exploratory mode* (see Figure 11) the user can choose a target variable and the system returns the list of variables that isolates it from the rest of the variables in the network (i.e. its Markov blanket) together with the measures (see Table 2) obtained by InSCo-Requisite. Then the posterior probability of the target variable, given the fixed evidences for the relevant variables, is obtained. In both modes, always is computed the posterior probability distribution of 'degree of revision' because it indicates us if the current SRS needs further revision.

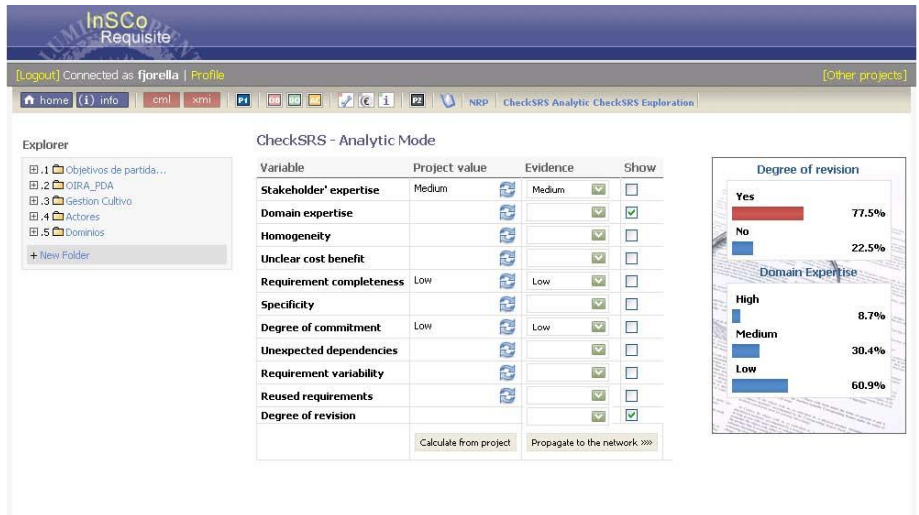

Figure 10: Analityc mode of InSCo-Requisite

Consider a situation in which the requirements engineer wants to check whether the current SRS can be considered as a baseline of a software project. In the analytic mode, the list of variables in Requisites is displayed and InSCo-Requisite can extract values of variables (displayed in column *Project value*) by applying the correspondence (see Table 2) between them and different metrics. The engineer can consider these values as evidences and can set them manually in the column *Evidence* (see Figure 10). The posterior probability of any target variable, given the assignment of evidences, can be obtained by clicking on *Propagate to the network*. In the case depicted in Figure 10, the probability value of 0.775 associated to the variable *degree of revision* indicates that the current SRS has to be reviewed.

Now, suppose that the requirements engineer is interested on 'requirements completeness'. The exploratory mode allows the engineer to choose this variable as the target. Automatically its set of neighbouring nodes (i.e. the variables that shield it from the rest of the network) are displayed in the column *Relevant variables*(see Figure 11). The evidence values of each one of them can be selected by means of drop-down list of values. Finally, the a posteriori probability of 'degree of revision', the target variable and all of its neigbours that have not receive evidence can be obtained by clicking on *Propagate to the network*.

The analytic mode lets find out what are the aspects of the project on which the developer should focus to improve the specification of requirements and to afford the next phases of software development with greater guarantees of success.

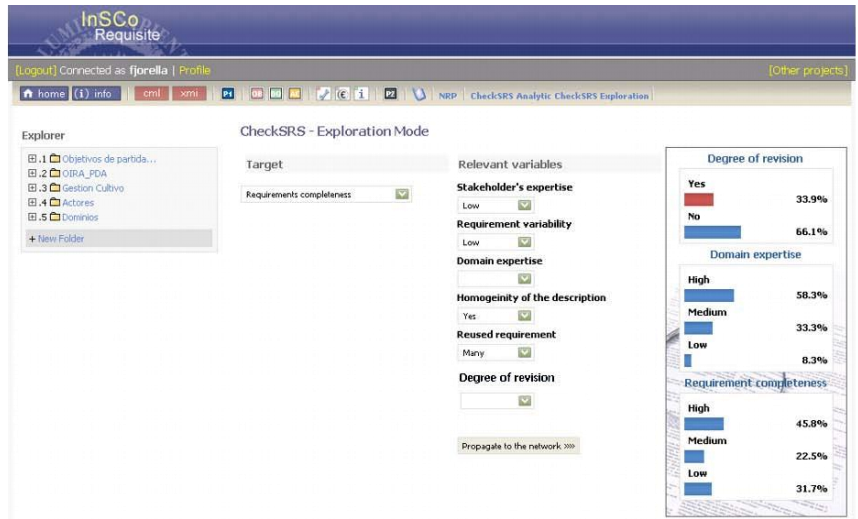

Figure 11: Exploratory mode of InsCo-Requisite

Thus, after setting the degree of revision, the developer can compare the values of the variables obtained by propagation in the Bayesian network, with the values obtained by direct measurement in the CARE tool. Based on this comparison, the developer will take corrective actions. Moreover, exploratory mode allows the engineer to focus on a single variable. In this way, the improvements to be made in project management can focus on actions that directly affect on the set of measures that influence this variable in order to keep it on a specific quality level.

# 5 Conclusions

Complex decision-making is a prominent aspect of Requirements Engineering since it is mainly a human activity that has the least technical load in the whole software project. In this work we have defined how to enhance the requirements specification development through the integration of a Bayesian network in a requirements management tool. The approach we have taken does not consist on solving a particular problem. Instead of it, we have instantiated an architecture that can be adapted to other models of reasoning and other software tools. This has been achieved by establishing a communication interface, between the academic CARE tool InSCo-Requisite and the Bayesian network Requisites, within a multilayer architecture. Besides we have proved the validity of Requisites not only by his integration on InSCo-Requisite, but also with the use of the evidences extracted from a large scale project, RALIC. The metrics calculated using this project data allow us to successfully predict the need of revision of the SRS.

The integration between Requisites and InSCo-Requisite provides the decision-maker a way of exploring the state of the work carried out on requirements using both analytic and exploratory model to improve the overal results of this stage.

In the next future, we plan to make an empirical evaluation of the enhanced InSCo-Requisite tool. We also want to refine the Bayesian network Requisites using the data recorded by this tool and to extend InSCo-Requisite functionalities to other requirements management tasks as traceability, change analysis and risk assessment.

## Acknowledgment

This research has been funded by the Spanish Ministry of Education, Culture and Sport under the projects TIN2010-20900-C04-02, TIN2015-71841-REDT.